\documentclass[10pt]{article}

\usepackage{fullpage}
\usepackage{setspace}
\usepackage{parskip}
\usepackage{titlesec}
\usepackage{placeins}
\usepackage{xcolor}
\usepackage{float}
\usepackage{lineno}
\usepackage{breqn}
\usepackage[toc,page]{appendix}
\usepackage{physics}
\usepackage{tikz}
\definecolor{fgreen}{RGB}{204,223,181}
\definecolor{arm}{RGB}{100,140,171}
\usetikzlibrary{arrows.meta, fit, shapes, positioning}

\usepackage{algorithm}
\usepackage{algpseudocode}

\usepackage[numbers,sort&compress]{natbib}

\usepackage{atbegshi}
\AtBeginDocument{\AtBeginShipoutNext{\AtBeginShipoutDiscard}}

\PassOptionsToPackage{hyphens}{url}
\usepackage[colorlinks = true,
            linkcolor = blue,
            urlcolor  = blue,
            citecolor = blue,
            anchorcolor = blue]{hyperref}
\usepackage{etoolbox}
\makeatletter
\patchcmd\@combinedblfloats{\box\@outputbox}{\unvbox\@outputbox}{}{%
}%
\makeatother

\renewenvironment{abstract}
  {{\bfseries\noindent{\large\abstractname}\par\nobreak}}

\titlespacing{\section}{0pt}{*3}{*1}
\titlespacing{\subsection}{0pt}{*2}{*0.5}
\titlespacing{\subsubsection}{0pt}{*1.5}{0pt}

\usepackage{authblk}

\usepackage{graphicx}
\usepackage[space]{grffile}
\usepackage{latexsym}
\usepackage{textcomp}
\usepackage{longtable}
\usepackage{tabulary}
\usepackage{booktabs,array,multirow}
\usepackage{amsfonts,amsmath,amssymb}
\providecommand\citet{\cite}
\providecommand\citep{\cite}

\newif\iflatexml\latexmlfalse
\DeclareGraphicsExtensions{.pdf,.PDF,.png,.PNG,.jpg,.JPG,.jpeg,.JPEG}

\usepackage[utf8]{inputenc}
\usepackage[english]{babel}
\DeclareUnicodeCharacter{0229}{e}

\begin{document}
\newcommand{\Sq}[1]{ \frac{ #1 }{ \sqrt{2} } }


\title{Entropy-Assisted Quality Pattern Identification in Finance}


\author[1]{Rishabh Gupta\textsuperscript{\ddag}}
\affil[1]{Department of Chemistry, Purdue University, West Lafayette, Indiana 47907, United States}
\author[2]{Shivam Gupta\textsuperscript{\ddag}}
\affil[2]{EntropyX Labs Pvt. Ltd., Ghaziabad, Uttar Pradesh, 201010, India}
\author[3]{Jaskirat Singh}
\affil[3]{Softure Solutions Pvt. Ltd., New Delhi, Delhi, 110059, India}
\author[1,4]{Sabre Kais}
\affil[4]{Department of Electrical and Computer Engineering, North Carolina State University, Raliegh, North Carolina, 27606, United States}
\thanks{skais@ncsu.edu}

\vspace{-1em}
  \date{\today}
\begingroup
\let\center\flushleft
\let\endcenter\endflushleft
\maketitle
\endgroup

\renewcommand{\thefootnote}{\ddag}
\footnotetext{These authors contributed equally to this work.}
\renewcommand{\thefootnote}{\arabic{footnote}}


\begin{abstract}
Short‐term patterns in financial time series form the cornerstone of many algorithmic trading strategies, yet extracting these patterns reliably from noisy market data remains a formidable challenge. In this paper, we propose an entropy‐assisted framework for identifying high‐quality, non‐overlapping patterns that exhibit consistent behavior over time. We ground our approach in the premise that historical patterns, when accurately clustered and pruned, can yield substantial predictive power for short‐term price movements. To achieve this, we incorporate an entropy‐based measure as a proxy for information gain: patterns that lead to high one‐sided movements in historical data, yet retain low local entropy, are more “informative” in signaling future market direction. Compared to conventional clustering techniques such as K-means and Gaussian Mixture Models (GMM), which often yield biased or unbalanced groupings, our approach emphasizes balance over a forced visual boundary, ensuring that quality patterns are not lost due to over-segmentation. By emphasizing both predictive purity (low local entropy) and historical profitability, our method achieves a balanced representation of Buy and Sell patterns, making it better suited for short‐term algorithmic trading strategies.

\end{abstract}

\maketitle

\section{Introduction}
In modern finance, the ability to recognize recurring short‐term patterns has become increasingly crucial for designing and executing algorithmic trading strategies. From high‐frequency trading desks to retail investors applying swing‐trading techniques, the recurring assumption is that historical price behavior repeats over time and the profitability of these techniques is directly proportional to the quality of these patterns. In fact, substantial volumes of historical data are routinely collected and mined to find these subtle yet exploitable patterns. However, the sheer level of noise in price series, driven by complex microstructure effects of the market and exogenous shocks, poses a constant obstacle to reliably distilling true signals from ephemeral artifacts. When attempting to exploit these signals in a systematic fashion, it becomes apparent that the success of any algorithmic model hinges on the quality of the underlying patterns.

A promising way to fortify pattern identification in such noisy contexts is to incorporate entropy as a measure of information content. Entropy is a versatile concept that has found applications in a wide range of fields, from information theory and statistical mechanics \cite{jaynes1957,rishabh2021,gupta2022variational,makhija2024time} to biology \cite{teschendorff2017single} and economics \cite{szczesniak2023,touchette2009large}. In finance, entropy is increasingly employed to quantify uncertainty in market behavior, assess risk, and enhance the robustness of trading models \cite{sensoy2019inefficiency}. Entropy, as defined in information theory, quantifies uncertainty or randomness within a probability distribution. For example, the Shannon entropy is defined as:
\begin{eqnarray}
H=-\sum_{i=1}^{N}p_{i}\ln{p_{i}},
\end{eqnarray}
where $p_i$ represents the probability of the occurrence of a particular event (or outcome). In the context of financial data, each 'event' could correspond to the future price movement following a specific short-term pattern in the data. When we apply this concept locally, looking at segments or groups within the historical feature space, we can measure how “pure” or consistent the outcomes are for similar patterns. A “pure” neighborhood is one where the outcomes are highly concentrated; for example, if nearly all occurrences of a given pattern lead to a large upward move, the local probability distribution is skewed toward that outcome and the resulting entropy is low. Thus, a low entropy value indicates that there is less uncertainty about what will happen next. Conversely, if a pattern is found in a region where outcomes are evenly split between upward and downward moves, the entropy is high, signaling greater uncertainty and suggesting that the pattern is less informative. 

This idea of leveraging entropy to derive inference in financial markets is not new \cite{benedetto2015maximum, ahn2019stock, zhou2013applications}. In our previous work, the EC-GBM (Entropy-Corrected Geometric Brownian Motion) \cite{gupta2024entropy}, we have demonstrated that by incorporating an entropy constraint into the model, one can effectively narrow down the forecast trajectories to those that reduce the overall uncertainty of the system. In the EC-GBM method, the predicted trajectories by the standard GBM \cite{stojkoski2019, stojkoski2020} are appended to the historical distribution, and the resulting change in entropy is computed. If a trajectory causes a significant drop in entropy relative to the reference state, it means that the trajectory improves the dominant features of the underlying distribution, effectively 'sharpening' the prediction by reducing uncertainty. The selected trajectories are then considered more reliable because they reflect a higher information gain or, equivalently, a more deterministic evolution of the price of the underlying asset. 

By integrating this entropy-based filtering mechanism into our pattern identification process, we not only eliminate overlapping or contradictory patterns but also preserve those that provide a clear, consistent signal. In our work, lower entropy is synonymous with higher informational content: patterns in low-entropy regions imply that the historical data exhibit a strong, unambiguous trend, which can be harnessed to improve the predictive power of algorithmic trading models. This stands in contrast to purely machine learning-based clustering methods, which could group patterns based solely on distance metrics without assessing their predictive clarity. In noisy financial markets, where overfitting and misclassification are constant threats, using entropy as an additional criterion ensures that our pattern selection process remains robust and focused on genuinely informative structures in the data. By maintaining only patterns that simultaneously exhibit low entropy and demonstrable historical profitability, we effectively increase the signal-to-noise ratio.

In summary, the use of entropy in our method enables us to quantify and enhance the quality of pattern identification. It acts as a natural filter that retains only those patterns that not only match well in a geometric sense, but also carry a high degree of predictive information, much like the entropy reduction principle demonstrated in EC-GBM for filtering out less relevant forecast trajectories. In this paper, we detail an end-to-end pipeline for entropy-assisted quality pattern identification and discuss how short‐term trading strategies can profit from explicitly integrating entropy measures. We also compare this approach to standard clustering‐based pattern detection, highlighting the role entropy plays in mitigating overfitting in volatile environments. Ultimately, we show that the focus of our method on non-overlapping high-information-gain patterns can lead to more reliable forecasts and improved performance in real‐time trading.

\begin{figure}
\centering
\includegraphics[scale=0.4]{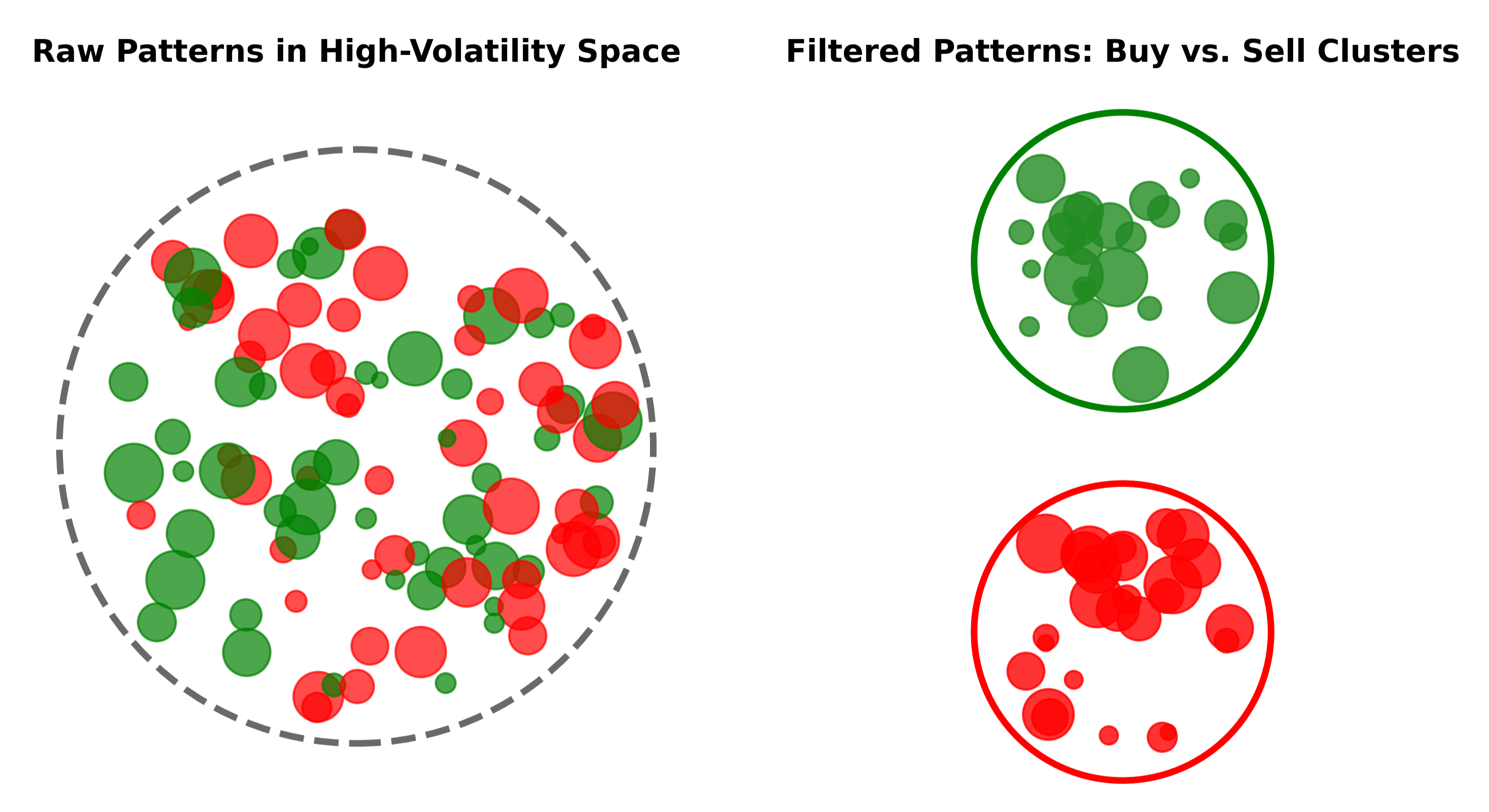}
\caption{The left panel illustrates raw trading patterns in a high‐volatility space, with each circle’s color indicating Buy (green) or Sell (red) and its size reflecting local entropy (smaller circles denote lower entropy and more consistent outcomes). These raw patterns exhibit extensive overlap and near‐duplicates across conflicting labels. In contrast, the right panel shows two non‐overlapping clusters following entropy‐based filtering, underscoring the method’s ability to isolate high‐quality, non‐ambiguous signals by removing contradictory patterns.} 
\label{fig01}
\end{figure}

\section{Methodology} \label{method}
The core objective of this work is to transform a large and noisy set of labeled short‐term trading patterns into two coherent clusters, one for Buy and one for Sell, such that no pattern in the Buy cluster overlaps with any pattern in the Sell cluster. As illustrated in Figure \ref{fig01}, raw patterns often exhibit significant overlap or near‐duplicates across conflicting labels, undermining their practical value in high‐volatility trading. We begin by collecting high‐resolution OHLC (open, high, low, close) data for a target asset (for example, Gold vs. USD) and identifying time segments that precede significant price swings (e.g., ±15 points within the subsequent two hours). Each extracted pattern is assigned a label (Buy or Sell) based on the direction of the ensuing market movement. This process results in a large pool of raw patterns, often numbering in the thousands, where many patterns overlap or nearly duplicate, even across opposing directions, which poses a challenge for robust signal extraction in algorithmic trading.

To address this challenge, we propose an entropy‐assisted filtering framework that operates in two main stages. First, we evaluate each pattern using a dual scoring system that combines a measure of local entropy with a historical profitability metric (PnL). The local entropy is computed by analyzing the immediate neighborhood of a given pattern in the high-dimensional feature space, derived from indicators such as H-L, C-O, H-O and O-L over the pattern window. If the majority of neighboring patterns share the same label, the local entropy is low, indicating a "pure" or unambiguous signal. Conversely, a high entropy value implies a mixed neighborhood, where the pattern’s outcome is less predictable. Simultaneously, we computed a PnL metric for each pattern, reflecting the average or maximum profit that could have been historically realized if a trade had been initiated when that pattern appeared. By normalizing the PnL values and combining them with the information gain (defined as the global entropy minus the local entropy) using a weighted sum, we obtain a final score for each pattern. This scoring mechanism ensures that the most valuable patterns are those that not only have high predictive certainty (low entropy), but also have demonstrated historical profitability.

In the second stage, we filter the raw patterns using a distance-based overlap criterion. Specifically, we define a distance threshold (based on the L1 (Manhattan) norm in the feature space) such that if a Buy pattern and a Sell pattern are found to be closer than this threshold, they are deemed to be overlapping and contradictory. In these cases, only the pattern with the higher combined score is retained. The end result is two refined sets—denoted 
B$'$ for Buy and S$'$ for Sell—that are non-overlapping across clusters yet may exhibit some controlled overlap within each cluster to capture the natural variation in similar trading scenarios. The core workflow of our entropy-assisted filtering method is presented in Algorithm \ref{alg:entropy}.

\begin{algorithm}[H]
\caption{Entropy-Assisted Quality Pattern Identification}
\label{alg:entropy}
\begin{algorithmic}[1]
\Require Raw pattern sets $\mathbf{B}$ (Buy) and $\mathbf{S}$ (Sell) extracted from OHLC data, distance threshold $\theta$, weight parameter $\alpha$, global entropy $H_{\text{global}}$
\Ensure Filtered Buy set $\mathbf{B}'$ and Sell set $\mathbf{S}'$ such that no pattern in $\mathbf{B}'$ overlaps with any pattern in $\mathbf{S}'$
  
\State $\mathbf{T} \gets \mathbf{B} \cup \mathbf{S}$
\ForAll{pattern $x \in \mathbf{T}$}
    \State Compute local entropy $H(x)$ based on the neighborhood in feature space
    \State Set information gain $\mathrm{IG}(x) \gets H_{\text{global}} - H(x)$
    \State Normalize historical profit/loss: $\mathrm{PnL}_{norm}(x)$
    \State Compute combined score: $\mathrm{score}(x) \gets \alpha \cdot \mathrm{IG}(x) + (1-\alpha) \cdot \mathrm{PnL}_{norm}(x)$
\EndFor
\State Sort $\mathbf{T}$ in descending order by $\mathrm{score}(x)$

\State Initialize $\mathbf{B}' \gets \emptyset$, $\mathbf{S}' \gets \emptyset$
\ForAll{pattern $x \in \mathbf{T}$ in sorted order}
    \If{$x$ is labeled \textbf{Buy}}
        \If{for every pattern $y \in \mathbf{S}'$, $d(x,y) \ge \theta$}
            \State $\mathbf{B}' \gets \mathbf{B}' \cup \{x\}$
        \EndIf
    \ElsIf{$x$ is labeled \textbf{Sell}}
        \If{for every pattern $y \in \mathbf{B}'$, $d(x,y) \ge \theta$}
            \State $\mathbf{S}' \gets \mathbf{S}' \cup \{x\}$
        \EndIf
    \EndIf
\EndFor
\State \Return $\mathbf{B}', \mathbf{S}'$
\end{algorithmic}
\end{algorithm}


\section{Results}
Our experiments were carried out on real-world data for Gold vs USD, which span from 2017 to 2023, with 2024 reserved exclusively for testing. A “pattern” in our study is defined as a eight-30 minutes segment of OHLC data that is represented by 32 features—specifically, eight values each for the differences H–L, C–O, H–O, and O–L—supplemented by a profitability (PnL) measure. By focusing on such short-term patterns extracted from raw historical data, we ensure that our methodology is tested under realistic market conditions that inherently include noise and irregularities not captured in simulated datasets. This realistic setting underscores the strength of our entropy-assisted filtering approach in distilling robust signals from noisy financial time series.
\begin{figure}
\centering
\includegraphics[scale=0.40]{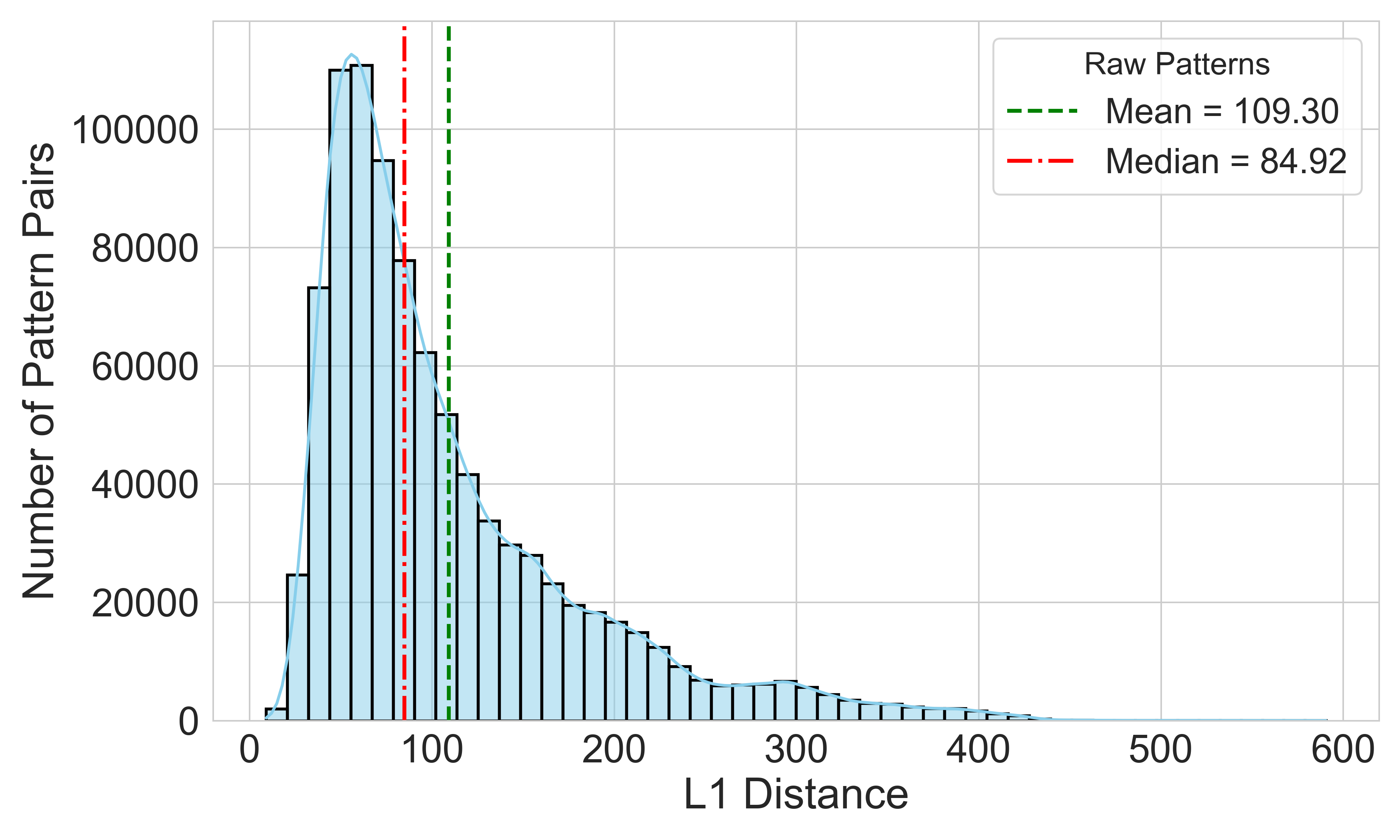}
\includegraphics[scale=0.40]{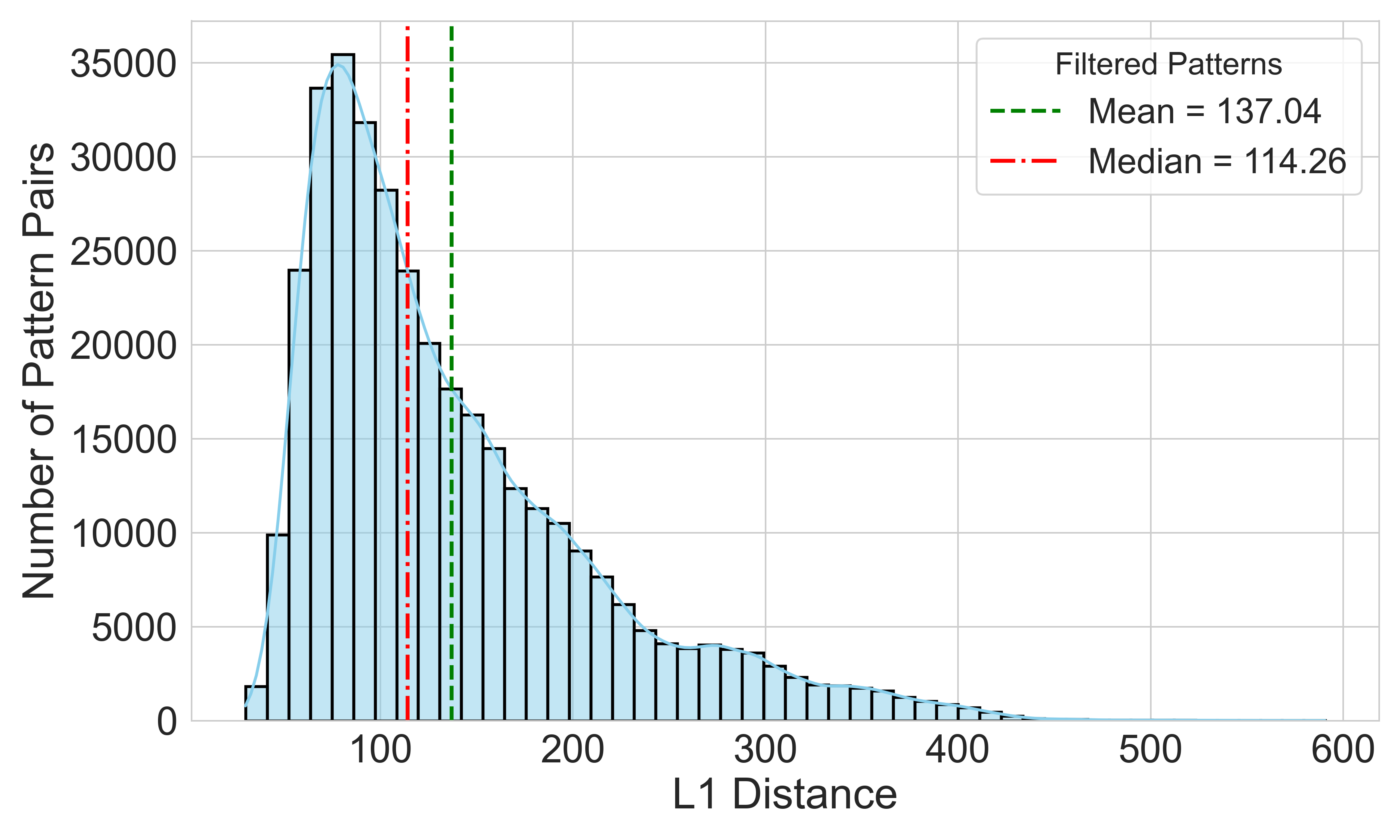}
\caption{The top histogram shows the distribution of pairwise L1 (Manhattan) distances between 900+ Buy and 1000+ Sell raw patterns. The bottom histogram presents the distances after entropy-based filtering, reducing the dataset to approximately 500 Buy and 600 Sell patterns. The significant reduction in pattern count is accompanied by an increase in mean and median distance, indicating that the filtering process effectively removes near-duplicates and conflicting patterns, ensuring that the remaining patterns are more distinct and non-overlapping in the feature space.} 
\label{fig02}
\end{figure}

Figure \ref{fig02} illustrates the effectiveness of our filtering process. The top histogram in Figure \ref{fig02} shows the distribution of pairwise L1 (Manhattan) distances between over 900 Buy and 1000 Sell raw patterns obtained from historical data that led to high volatility, revealing a substantial number of near-duplicates and overlapping instances. After applying our entropy-based filtering, which integrates local entropy (to assess pattern purity) and normalized historical profitability (PnL) into a combined score, the dataset is pruned to approximately 500 Buy and 600 Sell patterns. The bottom histogram in Figure \ref{fig02} demonstrates a notable increase in both the mean and median pairwise distances. This increase confirms that the filtering process effectively removes ambiguous and overlapping patterns, thereby retaining only those patterns that are truly distinct and non-overlapping in the feature space. The results for the filtered patterns in Figure \ref{fig02} are generated by keeping the alpha value at 0.8, thus giving more weight to the entropy factor compared to the PnL factor showcasing the relevance of entropy in this method. In this context, a 'quality' pattern is one that not only exhibits low local entropy, which implies high predictive consistency, but also delivers a strong historical PnL signal.
\begin{figure}
\centering
\includegraphics[scale=0.35]{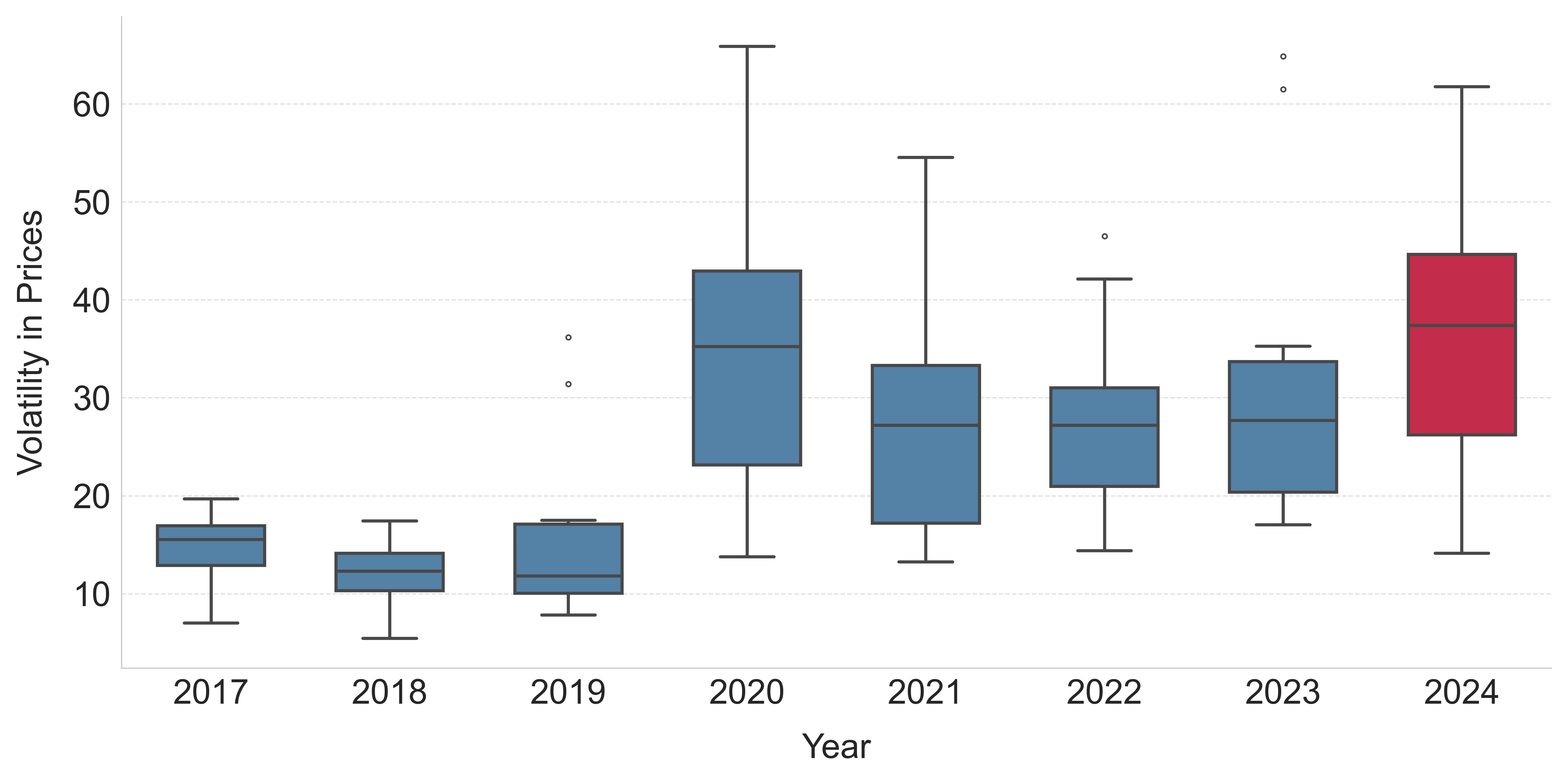}
\caption{Distribution of monthly volatility in gold prices from 2017 to 2024. The box plots depict the spread of monthly standard deviations of open prices within each year. Notably, the mean monthly volatility for 2024 is the highest among all years, indicating increased market fluctuations in the most recent period.} 
\label{fig03}
\end{figure}

\begin{figure}
\centering
\includegraphics[scale=0.35]{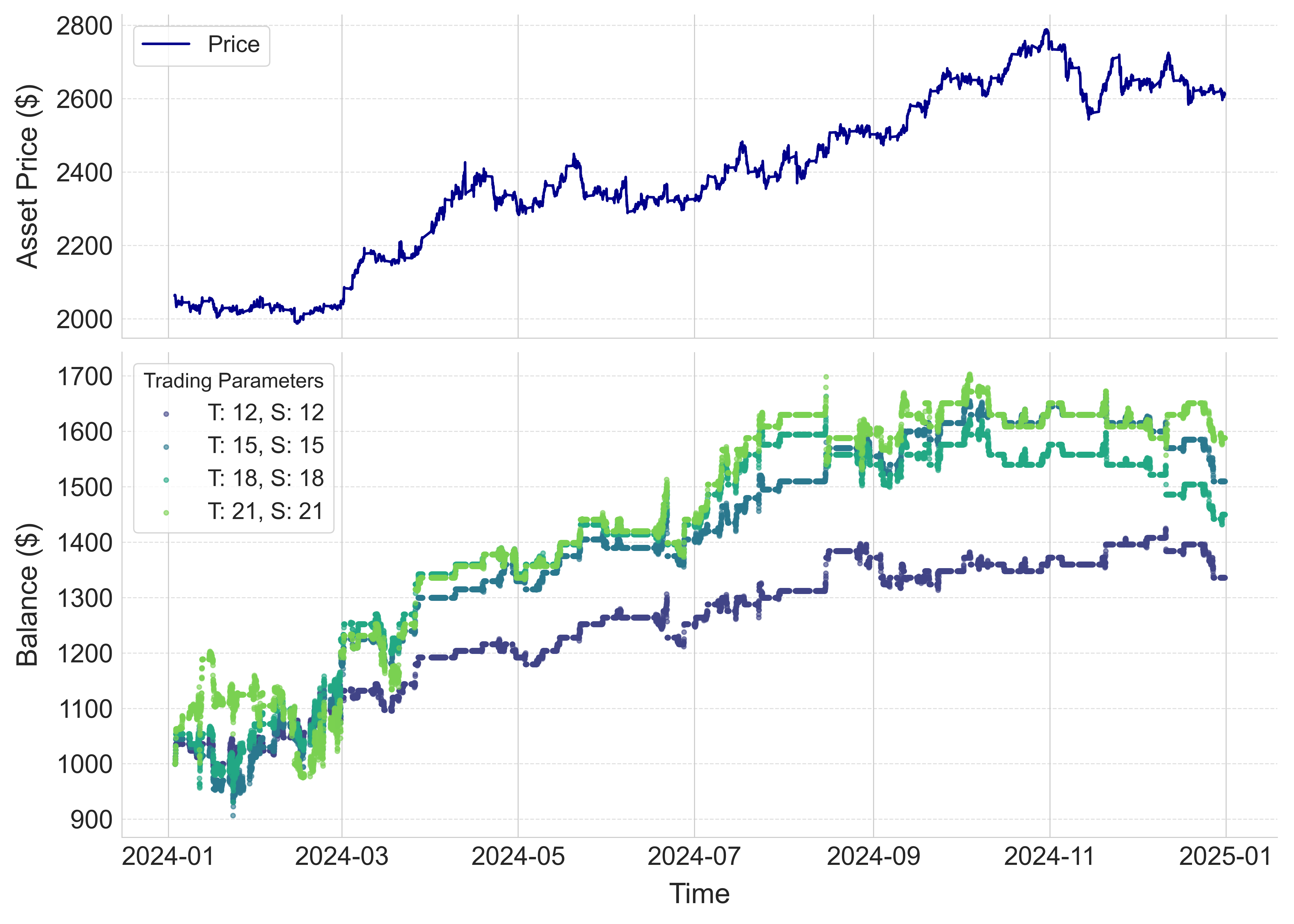}
\caption{The top subplot shows the asset's open price over time, while the bottom subplot tracks equity progression for different trading parameters (T: $target$, S: $stop loss$). The model consistently yielded profits across all configurations, demonstrating adaptability to market conditions and the impact of risk-reward trade-offs.} 
\label{fig04}
\end{figure}

Figure \ref{fig03} provides further context by depicting the monthly volatility distribution in gold prices from 2017 to 2024 using box plots of the standard deviation of open prices. Notably, 2024 exhibits the highest mean monthly volatility among all years. This escalation in volatility, particularly after the COVID period, reinforces the importance of short-term trading patterns. In an environment characterized by rapidly changing market sentiments and non-repeating long-term trends, short-term patterns serve as more reliable indicators of immediate market behavior. Their ability to capture transient market sentiments becomes even more critical as volatility increases, making the extraction of high-quality, non-overlapping patterns a vital component of robust algorithmic trading strategies. This point is further emphasized by the success of our algorithm in the real-world trading scenario, as depicted in Figure \ref{fig04}, which is discussed in the following.

Figure \ref{fig04} shows the practical performance of our trading strategy when applied to the unseen 2024 data. The upper sub-plot displays the evolution of the asset price over time, while the lower sub-plot tracks the progression of the investment under various configurations of the target and stop-loss parameters. Our back-testing framework is designed to trigger trades whenever a pattern match occurs: each order is executed with predefined target and stop loss values, a usual norm in algorithmic trading strategies. The model consistently generated profits across different parameter settings, resulting in annual returns ranging from 30$\%$ to 60$\%$. These results highlight not only the adaptability of our approach to varying market conditions but also the impact of carefully calibrated risk-reward trade-offs. In a real-time trading scenario, once a pattern match is identified, an order would be placed at the current open price, and the subsequent price action would be continuously monitored until either the target profit or the stop loss is reached, thereby ensuring disciplined exit strategies and robust performance.

\begin{figure}
\centering
\includegraphics[scale=0.35]{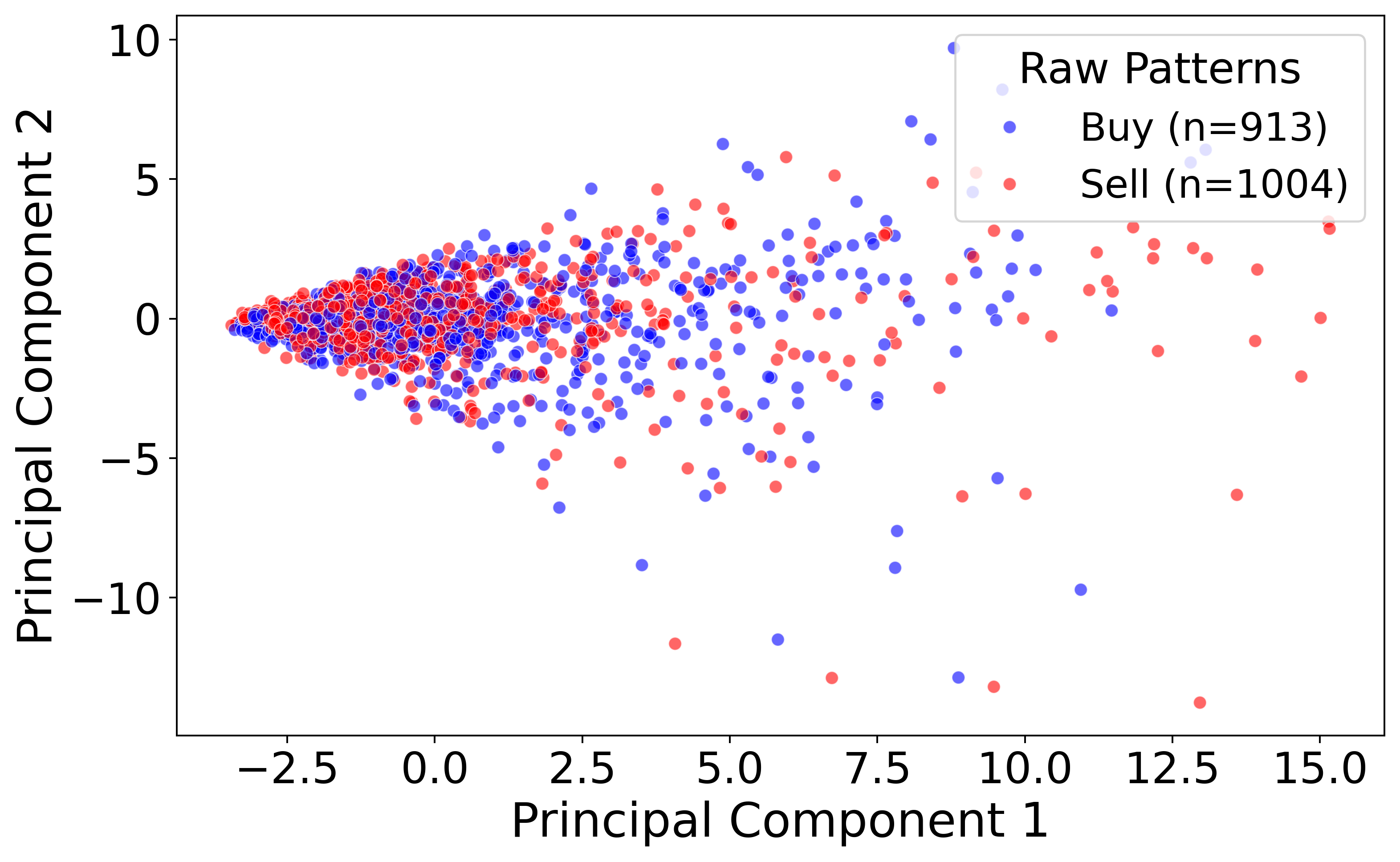}
\includegraphics[scale=0.35]{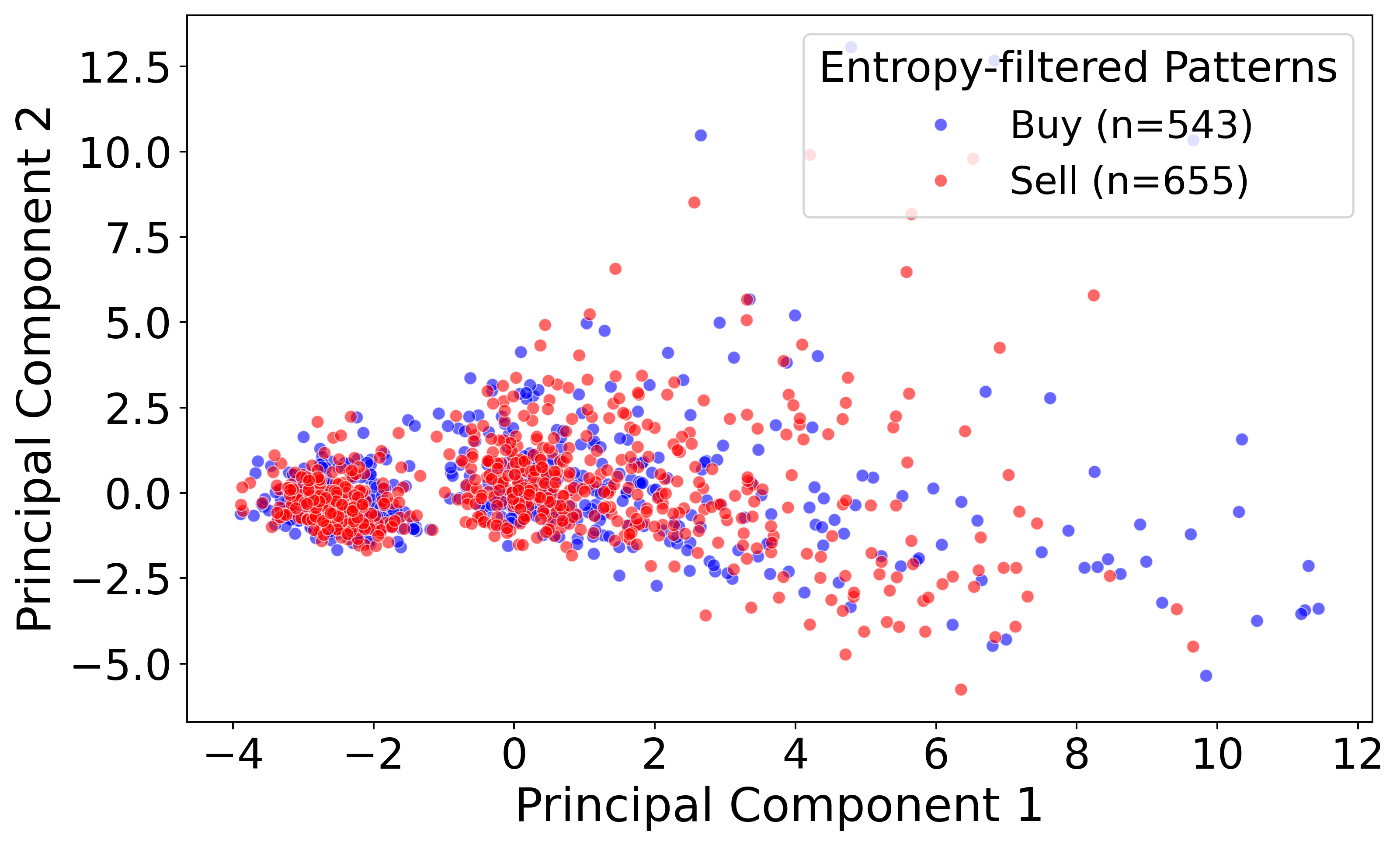}
\caption{Visualization of Buy (blue) and Sell (red) patterns in a two-dimensional PCA projection. The left subplot depicts the raw dataset (Buy = 913, Sell = 1004), which is highly intermixed in PCA space. The right subplot shows the entropy-filtered dataset (Buy = 543, Sell = 655), where ambiguous, overlapping patterns have been pruned. Although there is no crisp boundary in 2D, the filtering preserves a balanced Buy–Sell ratio similar to the raw data and removes contradictory patterns in the higher-dimensional feature space.} 
\label{fig05}
\end{figure}

\begin{figure}
\centering
\includegraphics[scale=0.35]{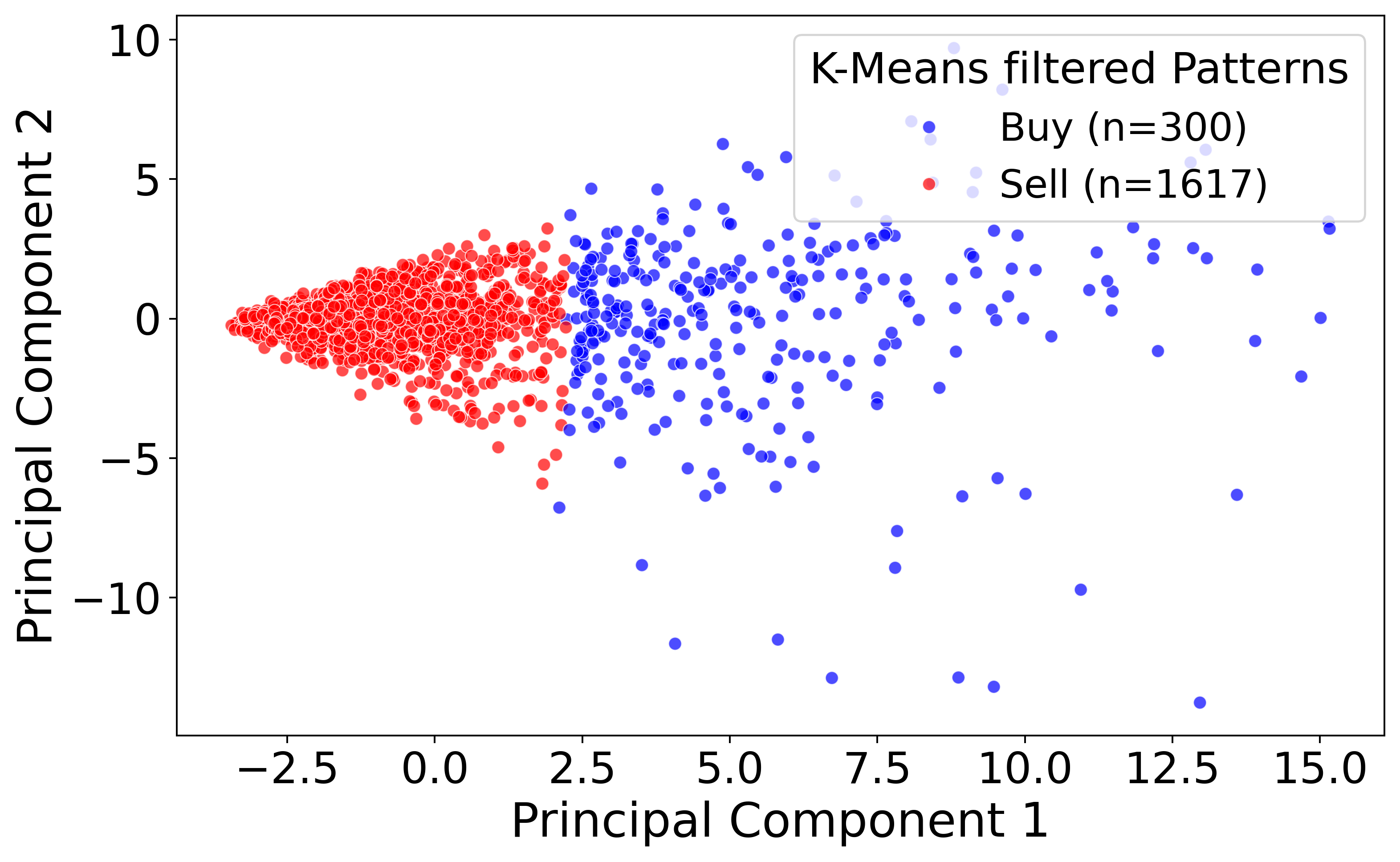}
\includegraphics[scale=0.35]{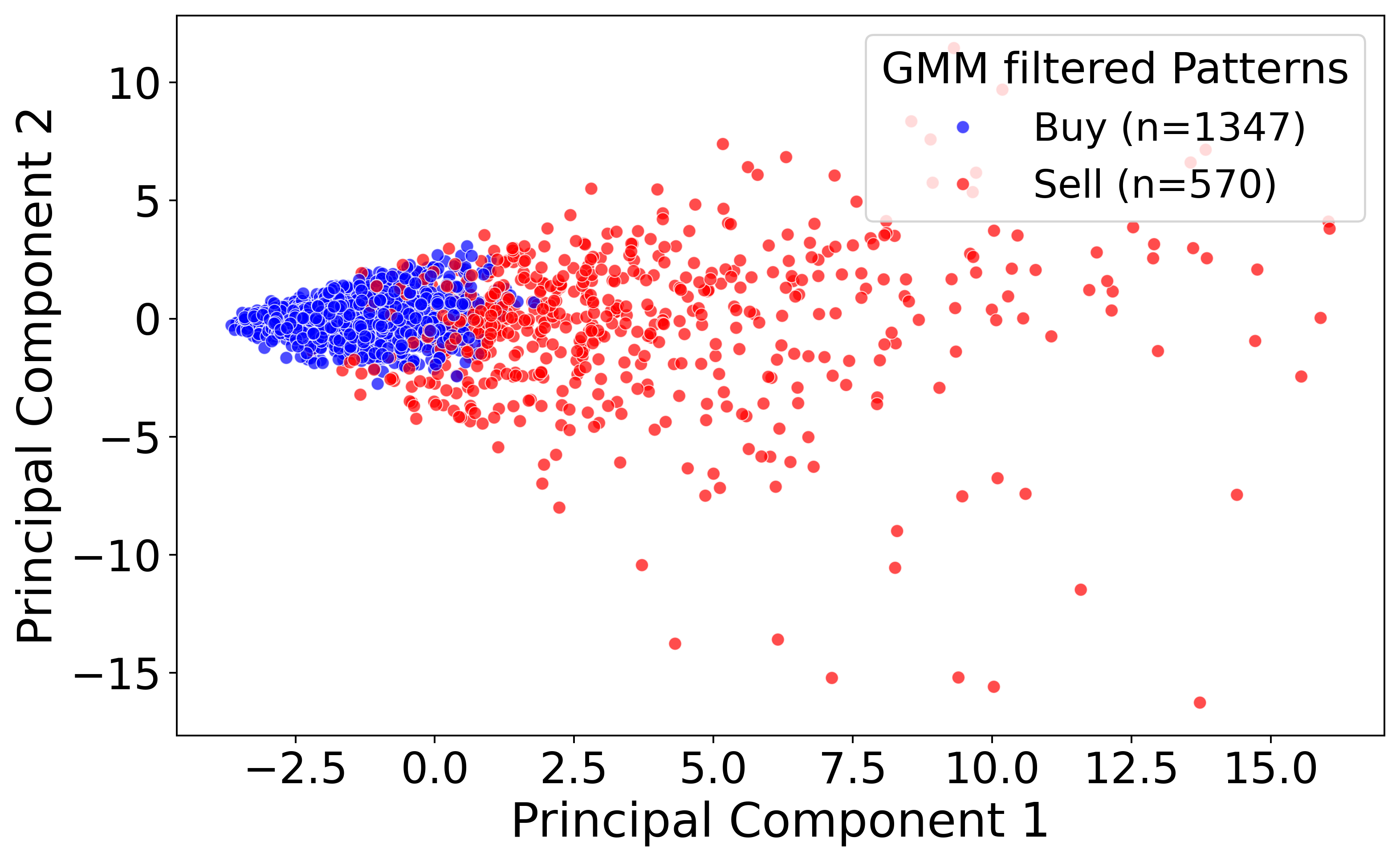}
\caption{Comparison of patterns produced by two standard clustering methods in a two-dimensional PCA projection. The left subplot illustrates K-means–filtered patterns (Buy = 300, Sell = 1617), and the right subplot shows GMM-filtered patterns (Buy = 1347, Sell = 570). While both algorithms yield visually distinct clusters, they produce heavily skewed Buy–Sell partitions. In contrast, the entropy-assisted approach (Figure \ref{fig05}) maintains a more balanced ratio of Buy vs. Sell patterns by explicitly accounting for historical profitability and directional consistency rather than relying solely on geometric proximity.} 
\label{fig06}
\end{figure}

This approach offers several advantages over traditional clustering techniques such as K-means \cite{he2007trading} or Gaussian Mixture Models (GMM) \cite{magazzino2022testing,ngoyi2023forex} which are two widely used clustering algorithms in financial market applications such as pattern identification, risk analysis, and anomaly detection. K-Means is a centroid-based algorithm that partitions data into a predetermined number of clusters by iteratively assigning each data point to its nearest centroid and then updating these centroids. However, K-Means assumes that clusters are spherical and well separated, which limits its effectiveness when the data exhibit significant overlap. In contrast, GMM adopts a probabilistic framework by modeling data as a mixture of multiple Gaussian distributions, thereby assigning each data point a probability of membership in each cluster. This soft clustering approach is better suited to financial scenarios where data distributions are complex and overlapping, such as in stock return modeling, risk-based portfolio optimization, and fraud detection. However, these standard clustering methods typically partition the data based solely on geometric distances, often resulting in imbalanced clusters (for instance, an excessively large Buy cluster and a very small Sell cluster) and fail to account for the directional consistency and historical profitability of the patterns. In Figure \ref{fig05}, we visualize the Buy vs. Sell patterns both before and after entropy‐assisted filtering using a two-dimensional principal component analysis (PCA) \cite{jolliffe2016principal}. Although the raw patterns appear heavily intermixed, the filtered set retains fewer, more distinctive patterns. Notably, the PCA projection does not exhibit a clear boundary separating Buy and Sell points. This outcome does not imply that the method fails to distinguish between the two sides in the higher‐dimensional feature space; rather, it reflects the inherent limitations of projecting complex financial data onto just two principal components. Figure \ref{fig06}, which compares K‐means and GMM results, illustrates that conventional clustering methods can indeed generate seemingly more distinct clusters in a 2D projection; however, these algorithms often yield disproportionately large clusters on one side, indicating potential bias, which is evident from the large number of patterns moved to one cluster in both of these clustering methods. Our approach avoids such pitfalls by enforcing non‐overlapping Buy and Sell sets without sacrificing a balanced representation of patterns. Consequently, while a crisp PCA boundary may not be visible, the underlying separation in the full feature space is preserved, leading to more reliable and equitable pattern sets for short‐term trading strategies. 

The entropy‐assisted method inherently prioritizes low-entropy patterns, i.e. those with a high level of outcome consistency - while also incorporating profitability metrics, thus producing a more balanced and informative set of signals. Furthermore, by explicitly removing overlapping patterns between Buy and Sell, our method avoids the confusion that can arise when ambiguous patterns are assigned to both clusters, leading to more robust and interpretable trading signals. Together, these results demonstrate that our entropy-assisted method not only yields a more balanced and high-quality pattern set compared to conventional clustering approaches (which tend to produce skewed clusters), but also translates into tangible trading performance improvements in a real, high-volatility market environment.

\section{Conclusion}

In this paper, we presented an entropy‐assisted framework that systematically transforms a raw, noisy set of short‐term trading patterns into two high-quality, non-overlapping clusters representing Buy and Sell signals. Our approach begins by extracting short-term patterns from high-resolution OHLC data, where each pattern—defined over a four-hour window—is represented by 32 engineered features along with an associated profit/loss (PnL) measure. By quantifying the local entropy of each pattern, we assess its predictive purity: patterns with low local entropy consistently lead to one directional move, while those with high entropy are ambiguous and less reliable. When combined with normalized PnL, this dual scoring mechanism enables us to retain only those patterns that are historically profitable and directionally consistent.

The effectiveness of our methodology is evidenced by the substantial reduction in the total number of patterns, from thousands of raw overlapping signals to a balanced set of approximately 500 Buy and 600 Sell patterns, while simultaneously increasing the average pairwise distance between patterns across clusters. This indicates that our filtering process effectively removes near-duplicates and conflicting signals. Unlike conventional clustering methods such as k-means or Gaussian Mixture Models, which often yield imbalanced or biased clusters due to their reliance on geometric proximity alone, our entropy-based approach emphasizes a balanced and interpretable representation of market signals. 

Looking ahead, future research could extend our framework in several promising directions. First, applying the methodology to a broader range of assets and markets would help validate its generalizability and robustness. Second, incorporating real-time adaptive mechanisms to update the pattern library based on incoming data may further improve the performance of algorithmic trading systems. Finally, exploring hybrid models that combine entropy-based filtering with machine learning approaches could yield even more refined predictive signals, opening up new avenues for risk management and portfolio optimization in complex and dynamic market environments.

Overall, our methodology provides a systematic, quantitative framework for transforming a raw, noisy set of short-term trading patterns into two high-quality, non-overlapping clusters that are both predictive and profitable. This dual filtering strategy, which combines entropy-based information gain with PnL normalization, ensures that the final pattern library is well suited for algorithmic trading applications, particularly in volatile market conditions where clarity and reliability of signals are paramount. Looking ahead, our framework offers promising avenues for future research. In particular, we propose exploring the substitution of Shannon entropy with Tsallis entropy as an alternative measure of uncertainty. Tsallis entropy, with its adjustable parameter that can tune the sensitivity to rare events, may offer enhanced flexibility in capturing the complex, multifractal nature of financial time series, potentially leading to further improvements in pattern quality and trading performance. Moreover, drawing inspiration from global optimization techniques, such as the pivot methods \cite{stanton1997new,serra1997pivot,serra1997comparison,serra1997symmetry,nigra1999pivot,nigra2001study}, future work can explore pivot moves through phase space guided by a q-distribution based on Tsallis entropy. This hybrid approach could enhance our ability to identify and relocate patterns within the feature space, potentially leading to even more efficient and adaptive algorithmic trading strategies.
\section*{Data availability}
The datasets generated during the current study are available from the corresponding author on reasonable request, while the financial datasets employed are publicly available online at www.histdata.com.
\section*{Acknowledgements}
S.K.  would also like to acknowledge funding from the U.S. Department of Energy (DOE) (Office of Basic Energy Sciences), under Award No. DE-SC0019215.

\bibliographystyle{apsrev}
\bibliography{bibliography}

\end{document}